\documentclass[preprint,prd,nofootinbib]{revtex4}

\usepackage{color} 
     
\usepackage{amssymb}
\usepackage{amsmath}
\usepackage{graphicx}
     
\newcommand{\pderiv}[2]{\frac{\partial #1}{\partial #2}}

\begin{document}

\title{Schwinger effect for non-Abelian gauge bosons}

\author{Michael Ragsdale}
\email{raggy65@mail.fresnostate.edu}
\affiliation{Department of Physics, California State University Fresno, Fresno, CA 93740-8031, USA}

\author{Douglas Singleton}
\email{dougs@csufresno.edu}
\affiliation{Department of Physics, California State University Fresno, Fresno, CA 93740-8031, USA \\
and \\
Institute of Experimental and Theoretical Physics Al-Farabi KazNU, Almaty, 050040, Kazakhstan}

\date{\today}

\begin{abstract}
We investigate the Schwinger effect for the gauge bosons in an unbroken non-Abelian gauge theory ({\it e.g.} the gluons
of QCD). We consider both constant``color electric" fields and ``color magnetic" fields as backgrounds. As in the Abelian Schwinger effect
we find there is production of ``gluons" for the color electric field, but no particle production for the color magnetic field case. Since 
the non-Abelian gauge bosons are massless there is no exponential suppression of particle production due to the mass of the electron/positron 
that one finds in the Abelian Schwinger effect. Despite the lack of an exponential suppression of the gluon production rate due to the masslessness
of the gluons, we find that the critical field strength is even larger in the non-Abelian case as compared to the Abelian case. This is the result
of the confinement phenomenon on QCD.  
\end{abstract}

\maketitle

\section{Introduction}

The Schwinger effect \cite{schwinger, heisenberg} is the creation of electron-positron pairs from a uniform electric field. The energy to create the pairs
comes from energy of the electric field. Since the rest mass energy of the electron is large relative to the electric field energy density that one 
can achieve in the laboratory the Schwinger effect has not been observed experimentally in the form in which it was first calculated -- a uniform background
electric field producing electron-positron pairs. The reason for this is the probability per unit volume per unit time of creating $e^+ e^-$ pairs
is given by 
\begin{equation}
\label{ee}
\frac{{\rm prob}_{e^+e^-}}{{\rm Vol} \times {\rm time}} \propto e^2 E^2 _{EM} \exp \left( - \frac{\pi m^2}{e E_{EM}} \right) ~, 
\end{equation}     
where $e$ and $m$ are the charge and mass of the electron/positron, $E_{EM}$ is the magnitude of the uniform electric field and $c$ and $\hbar$ have been set to 1. Since the electron/positron have a finite rest mass the exponential will suppress $e^+ e^-$ production unless $\frac{\pi m^2}{e E_{EM}} \sim 1$. If one takes $\frac{\pi m^2}{e E_{EM}} = 1$, restores factors of $c$ and $\hbar$ then one finds the this uniform electric field magnitude is enormous -- $E_{EM} = 1.4 \times 10^{14} \frac{dyne}{esu}$ or $E_{EM} = 4.2 \times 10^{18} \frac{N}{C}$ -- well beyond the present ability to create in a laboratory. If electron/positrons were lighter, or massless, one would more readily be able to observe the electromagnetic Schwinger effect. 

While the electron/positron mass is non-zero there is a system where a charged particle is massless and thus the Schwinger effect should not have the exponential suppression found in the electromagnetic case. The is the case of gluons in quantum chromodynamics (QCD). Gluons are massless and carry color charge due to the non-Abelian nature of QCD. Thus we want to investigate the QCD version of the Schwinger effect whereby a constant background color ``electric" field creates gluons. Of course due to color confinement 
one can not really make a constant color field over a macroscopic distance. However within QCD bound systems one might think of the quantum chromodynamic flux tubes that are postulated to bind quarks together into mesons/baryons as giving a uniform color electric and color magnetic fields inside the tubes. 

In this work we consider, for simplicity, the $SU(2)$ non-Abelian gauge theory. This has almost all of the features of larger non-Abelian groups like $SU(3)$ but the details of
the calculations are simpler and more transparent. Often in this work we will mention ``gluons" or QCD which technically refer to the strong $SU(3)$ gauge group, but we mean by this
the $SU(2)$ toy model of the true $SU(3)$ interaction.       

\section{Brief review of electric Schwinger effect}

In this section we will give a very brief overview of Schwinger's method for calculating pair production in the case of Abelian electromagnetic fields. In the following
section we will use this background to address pair production in a non-Abelian color electric and color magnetic fields. In addition to the original articles on the Schwinger effect
\cite{schwinger, heisenberg} there are many good discussion of this method in the literature. A very small sample of these are found in references \cite{holstein} \cite{paddy} \cite{brout}
\cite{gelis}. We will follow most closely the pedagogical article \cite{holstein}. 

To begin the vacuum to vacuum transition amplitude is given very generally by the expression
\begin{equation}
\label{vac-vac}
{\rm amp(vac \rightarrow vac)} \longrightarrow \int [d\phi] \exp \left( i \int d^4x {\cal L} (\phi, \partial _\mu \phi ) \right) ~,
\end{equation}
where $\phi$ is some generic field, $\int [d\phi]$ is a path integral over different field configurations, and ${\cal L} (\phi, \partial _\mu \phi )$ is the
Lagrange density for the field $\phi$. The Lagrange density will lead to an equation of motion for the field $\phi(x)$ in terms of some operator ${\cal O} \phi$ \footnote{In the example used in Holstein \cite{holstein} of a scalar field of mass $m$ and charge $e$ the operator is
${\cal O} = (\partial _\mu + i e A_\mu(x) )^2 + m^2$ where $A_\mu$ is the electromagnetic vector potential of the electromagnetic field.}. The field $\phi(x)$ can be expanded as $\phi(x) = \sum _n a_n \chi _n (x)$ where $\chi_n (x)$ are eigenstates of the equation
of motion given by 
\begin{equation}
\label{eigen}
{\cal O} \chi _n (x) = \lambda _n \chi _n (x) ~,
\end{equation}   
with $\lambda _n$ being the eigenvalues. In terms of the operator ${\cal O}$ the vacuum to vacuum amplitude is
 ${\rm amp(vac \rightarrow vac)} \sim \frac{const.}{det {\cal O}}$ where the determinant of the operator can be
written in terms the eigenvalues as $det {\cal O} = \prod _n \lambda _n$. Using all this we can write out the 
results as
\begin{equation}
\label{deto}
\frac{1}{det {\cal O}} = \exp[- \ln (det {\cal O})] = \exp \left[ -\ln \prod _n \lambda _n \right] =
\exp\left[ - \sum _n \ln \lambda _n \right] = \exp[- tr \ln ({\cal O} ) ] ~.
\end{equation}
Next we use the representation of the logarithm as $\ln \lambda _n = - \int _0 ^\infty \frac{\exp (-\lambda _n s)}{s} ds$ to
write the middle expression in (\ref{deto}) as
\begin{equation}
\label{vac-vac2}
tr (\ln {\cal O}) = - \sum _n \int _0 ^\infty \frac{\exp (-\lambda _n s)}{s} ds \equiv \zeta ~.
\end{equation}
Finally the vacuum to vacuum amplitude from (\ref{vac-vac}) becomes
\begin{equation}
\label{vac-vac3}
{\rm amp(vac \rightarrow vac)}  \propto \exp[- \zeta ] ~.
\end{equation}
If $\zeta$ has a real part ({\it i.e.} $\gamma =Re (\zeta)$) then one can square the amplitude to get the 
probability for pair production as
\begin{equation}
\label{prob} 
{\rm prob}_{pair} = (1 - \exp[-2 \gamma]) ~.
\end{equation}

This is the basic procedure which we will apply to the pair production of gluons in uniform color electric and color 
magnetic fields. However before moving on to this we give a few more details about the pair production of scalar particles 
of mass $m$ and charge $e$ in a uniform electric field. For a uniform electric field in the $z$-direction 
${\bf E} = E_0 {\hat z}$ the vector potential can be of the form ${\bf A} (t) = - E_0 t {\hat z}$ or
$\phi (z) = - E_0 z$. (In the QCD case we will find a similar situation for the QCD potentials). For the time-dependent
vector potential, ${\bf A} (t) = - E_0 t {\hat z}$, the operator in footnote 1 becomes
\begin{equation}
\label{oper}
{\cal O} = \partial _t ^2 - (\partial_z - i e E_0 t )^2 - \partial _x ^2 - \partial _y ^2 + m^2 ~.
\end{equation}   
The eigenvalues connected with (\ref{oper}) are
\begin{equation}
\label{eigen-2}
\lambda _n = e E_0 (2 n +1) + p_x^2 + p_y^2 + m^2 ~,~~~~ {\rm n =0, 1, 2...}~.
\end{equation}  
Using this eigenvalues in (\ref{vac-vac2}) and performing the sum \footnote{It is only the discrete part of $\lambda_n$ ({\it i.e.} $e E_0 (2 n +1)$ which is summed over. The ``sum" over the momenta part of $\lambda_n$ ({\it i.e.} $p_x, p_y$ and $p_z$) involve integrations. The details
leading to the result in (\ref{zeta}) are in reference \cite{holstein}, and in addition we perform essentially the same steps in appendix A when we calculate the production of gluons in a uniform chromoelectric field.} of the different pieces of $\lambda _n$ 
yields 
\begin{equation}
\label{zeta}
\zeta = - i L^3 T \frac{e E_{EM}}{16 \pi^2} \int _0 ^\infty \frac{ds}{s^2} \frac{\exp[-m^2 s]}{\sin(e E_{EM} s)} ~,
\end{equation}
where $L$ and $T$ are the spatial and temporal size to ``cube" inside which the system is quantized. The factor of $i$ comes from doing a rotation to imaginary time and back ({\it i.e.} $t \rightarrow i T$). This time rotation also involves the change $E_{EM} \rightarrow i E_{EM}$. We will use the same procedure when we carry out the QCD version of this calculation. In order to have particle production $\zeta$ 
needs to have a real part which will occur if the integral in (\ref{zeta}) has an imaginary part. The integral in (\ref{zeta}) does
have imaginary parts coming from the contour integrations involving the poles in the integrand at $s_n = n \pi / e E_{EM}$. (The $1/s^2$ divergence in the integrand in (\ref{zeta}) leads to an infinite imaginary part which is removede via renormalization \cite{holstein2}).
Taking semi-circular deviations around each of the poles $s_n$ leads to a real part of $\zeta$ given by
\begin{eqnarray}
\label{gamma}
\gamma = Re(\zeta) &=& L^3 T \frac{e^2 E_{EM} ^2}{16 \pi^3} \sum_{n=1} ^\infty \frac{(-1)^{n+1}}{n^2} \exp 
\left(- \frac{\pi n m^2}{e E_{EM}} \right) \nonumber \\
&\approx& L^3 T \frac{e^2 E_{EM} ^2}{16 \pi^3} \exp  \left(- \frac{\pi m^2}{e E_{EM}} \right)~.
\end{eqnarray}
Due to the exponential term the main contribution comes from $n=1$ term in the sum. The result in (\ref{gamma}) can be used to 
obtain the result in (\ref{ee}) by inserting it into (\ref{prob}) and dividing by (Vol $\times$ time)
\begin{eqnarray}
\label{ee1}
\frac{{\rm prob}_{e^+e^-}}{{\rm Vol} \times {\rm time}} &=&  \frac{1}{{\rm Vol} \times {\rm time}} (1-e^{-2 \gamma})
\approx \frac{e^2 E^2 _{EM}}{8 \pi ^3 \hbar ^2 c} \exp \left( - \frac{\pi m^2 c^3}{e E_{EM} \hbar} \right) \nonumber \\
&=& \frac{\alpha_{EM} E^2 _{EM}}{8 \pi ^3 \hbar} \exp \left( - \frac{\pi m^2 c^3}{e E_{EM} \hbar} \right)~, 
\end{eqnarray} 
where we have restored factors of $\hbar$ and $c$ and then written the result in terms of the fine structure constant
$\alpha _{EM} = \frac{e^2}{\hbar c} \approx \frac{1}{137}$. We want to use (\ref{ee1}) to determine ${\rm prob}_{e^+e^-}$
when the electric field is at the critical value $E_{EM} \approx 1.4 \times 10^{14} \frac{dyne}{esu}$. To do this we need some
way to characterize what a ``natural" value is for the volume and time in the denominator of (\ref{ee1}). For the volume we take
it to be the cube of the reduced Compton wave length of the electron ${\rm Vol} = \left( \frac{\hbar}{mc} \right) ^3 
\approx 5.7\times 10^{-32} cm ^3$. For the time we take the reduced Compton time ${\rm time} = \frac{\hbar}{mc^2} 
\approx 1.3 \times 10^{-21} sec$. Putting all these values together in (\ref{ee1}) yields
\begin{equation}
\label{ee2}
{\rm prob}_{e^+e^-} \approx 0.015
\end{equation} 

In the next section we will calculate the equivalent result to equation (\ref{ee1}) but for gluons produced from uniform chromoelectric fields
and uniform chromomagnetic fields. We will set the probability of gluon production ({\it i.e.} ${\rm prob}_{gg}$) to the same 
value as that of $e^+ e^-$ production from (\ref{ee2}) for the critical electric field, $E_{EM}$. In this way we will arrive at
the critical chromoelectric/chromomagnetic field strength to produce gluons. The initial expectation might be that the critical 
color field strength might be smaller since gluons are massless so there will be no exponential suppression as occurs due to the
rest mass of the electron for the QED case.   

\section{Schwinger effect for uniform color electric field and uniform color magnetic field} 

To calculate the production rate for gluons from uniform chromoelectric and chromomagnetic fields we follow reference
\cite{greiner} and place gluon field excitations in a fixed background field. For the two background fields we take 
a uniform chromoelectric field and uniform chromomagnetic field. These calculations are rather involved and we put the 
details in appendix A for the uniform chromoelectric field and appendix B for the uniform chromomagnetic field. The result 
for the production rate for gluons in a uniform chromoelectric field is given by equation (\ref{gamma-4}) as 
\begin{equation}
\label{gamma-1}
\gamma \approx 0.00187 \times L^3T\frac{g^2E_0^2}{16\pi^3} ~.
\end{equation}
Using this result from (\ref{gamma-1}) in (\ref{prob}) we obtain the probability for SU(2) gluon creation per unit volume and unit time
\begin{equation}
\label{gg}
\frac{{\rm prob}_{gg}}{{\rm Vol} \times {\rm time}} =  \frac{1}{{\rm Vol} \times {\rm time}} (1-e^{-2 \gamma})
\approx \frac{0.00187 g^2 E_0^2}{8 \pi^3 \hbar^2  c} = \frac{0.00187 \alpha_{QCD} E_0^2}{8 \pi^3 \hbar }~.
\end{equation}
In the last step we have restored factors of $\hbar$ and $c$ and introduced the QCD fine structure constant 
$\alpha _{QCD} =\frac{g^2}{\hbar c}$. We will take $\alpha _{QCD} \approx 1$ so that we are in non-perturbative regime. For the present case
we take the distance scale to be $10^{-12} cm$. This is one order of magnitude larger than the typical strong interaction bound state size
of $1 {\rm fermi}$. Thus is the volume factor in (\ref{gg}) we have ${\rm Vol} = 10^{-36} cm^3$ and for the time factor we take
${\rm time} = \frac{10^{-12}}{3 \times 10^{10}} = 3.3 \times 10^{-23} {\rm sec}$. Finally we take the value for the probability of gluon production,
${\rm prob}_{gg}$ at the critical chromoelectric field magnitude to be the same as that for the electromagnetic case as given in 
(\ref{ee2}) namely ${\rm prob_{gg}} \approx 0.015$, using these assumptions in (\ref{gg}) we can calculate the value of the critical 
chromoelectric field magnitude as
\begin{equation}
\label{eqcd-cr}
E_0 \approx 2.5 \times 10^{17} \frac{\rm dyne}{\rm esu}~.
\end{equation}
This is then three orders of magnitude larger than the $E\&M$ critical field value of $E_{EM} \approx 1.4 \times 10^{14} \frac{dyne}{esu}$. 
Thus although gluons are massless (and thus there is no exponential suppression factor as in the electromagnetic case due to 
the mass of the electron) this does not lower the critical chromoelectric field value, and in fact we find the critical field value for
QCD is three orders of magnitude larger than in the $E\&M$ case. This, at first sight, surprising result arises from the fact that QCD has
a confinement scale in the range of $1 {\rm fermi} =10^{-13} {\rm cm}$. This distance is smaller than the Compton wave length of the electron 
which we used to set the volume and time in the electromagnetic case as ${\rm Vol} = \left( \frac{\hbar}{mc} \right) ^3 
\approx 5.7 \times 10^{-32} cm ^3$ and ${\rm time} = \frac{\hbar}{mc^2} \approx 1.3 \times 10^{-21} sec$ respectively. In the QCD case
we overestimated the distance scale as 10 times larger than $1 {\rm fermi} =10^{-13} {\rm cm}$ and also used this to obtain the characteristic time
for the QCD case. Even though we (slightly) overestimated the QCD distance and time scales (which would by (\ref{gg}) would be favorable toward to decreasing 
the critical QCD field strength) we nevertheless found that the critical QCD field strength was larger than the critical $E\&M$ critical field strength. 
Even if we had used (without justification) the electron Compton wave length and associated Compton time in (\ref{gg}) this would give a QCD
critical field strength of $E_0 \sim 10^{14} \frac{dyne}{esu}$ {\it i.e.} the same as for $E\&M$. Thus the masslessness of the gluons does not
lead to a lower critical field strength relative to what is found in the canonical Schwinger mechanism case.

\section{Conclusion}

We presented the Schwinger effect for SU(2) gluons. We did this since gluons being massless we expect this effect to be more important than standard Schwinger 
effect which is exponentially suppressed due to electron/positron rest mass. The calculation details for gluon production from a uniform 
chromoelectric field and from a uniform chromomagnetic field are found in appendix A and B respectively. As in the electromagnetic case a chromoelectric field
will produce gluons while the chromomagnetic field does not.  

The motivation for investigating gluon production, as opposed to the production of quarks, was that since gluons are massless there will not be an
exponential suppression due to the mass of the quarks that one finds in the electromagnetic case due to the mass of the electron ({\it i.e.} the
last exponential term in (\ref{ee}). However, due to confinement the natural length scales and time scale for the QCD case were several orders of 
magnitude smaller than the similar scales for the electromagnetic case which were set to the reduced Compton length of the electron and the associated
time. This in turn made the factor ${\rm Vol \times time}$ smaller for the QCD case as compared to the $E\&M$ case which in turn, given the expression for
the probability for production per unit volume per unit time in (\ref{gg}) gave a larger magnitude for the chromoelectric field for producing gluons
as compared to the magnitude of the electric field for producing electrons and positrons.

{\bf Acknowledgment}

DS is supported by grant $\Phi.0755$ in fundamental research in Natural Sciences by the Ministry of Education and Science of Kazakhstan.

\appendix

\section{Uniform Color Electric Field Calculation}

\setcounter{section}{1}

To have a constant color electric field in the $\hat{z} = 3$ spatial direction and in the $a=3$ color direction, 
\begin{equation}
\label{eqcd}
E^a _\mu \rightarrow E^{3} _{3} = F^3 _{03} = \partial _0 A_3 ^3 - \partial _3 A_0 ^3 +g \epsilon^{3bc} A^b_3 A^c_0 = E_0 \hat{z} ~,
\end{equation}
there are two gauge choices one can make for the potential 
\begin{equation}
\label{aqcd}
(i) ~~ A_\mu^{a} = E_0 t \delta_{\mu 3} \delta^{a3}  ~~~~~;~~~~~ (ii) ~~ A_\mu^{a} = -E_0 z \delta_{\mu 0} \delta^{a3}
\end{equation}

For the calculation in this appendix we use the $(i)$ form of the potential. Inserting form $(i)$ from
(\ref{aqcd}) into (\ref{eqcd}) does give  $E^a _\mu \rightarrow E^3 _3 = \partial _0 A_3 ^3  = E_0 \hat{z}$. We now take
the potential $(i)$ from (\ref{aqcd}) as a background potential ({\it i.e.} $A_\mu^{(0)a}$) and we consider small
variation $A_\nu^a$ around this background (the background nature of the potential is indicated by the superscript
$(0)$). In this way the QCD action can be written as
\begin{equation}	
\label{ff1}
	F_{\mu\nu}^a F^{a\mu\nu}=\left[\left(\partial_\mu A_\nu^{(0)a}-\partial_\nu A_\mu^{(0)a}\right)+\left(\partial_\mu A_\nu^a-\partial_\nu A_\mu^a\right)+g\epsilon^{abc}\left(A_\mu^{(0)b}+A_\mu^b\right)\left(A_\nu^{(0)c}+A_\nu^c\right)\right]^2
\end{equation}	
We now run through the color indices $a=1,2,3$ and insert the explcit form of the background potential $A_\mu^{(0)3}=E_0 t\delta_{\mu 3}$. The only non-zero values of $\partial_\mu A_\nu^{(0)3}-\partial_\nu A_\mu^{(0)3} +g \epsilon^{3bc} A^{(0)b}_\mu A^{(0)c}_\nu$ are when $\mu=0$ and 
$\nu=3$ or vice versa, each of which contribute a term of $E_0$.
\begin{eqnarray}
F_{\mu\nu}^a F^{a\mu\nu}&=&\left\{\partial_\mu A_\nu^1-\partial_\nu A_\mu^1+g\left[A_\mu^2\left(A_\nu^3+E_0 t\delta_{\nu 3}\right) - \left(A_\mu^3+E_0 t\delta_{\mu 3}\right)A_\nu^2\right]\right\}^2 \nonumber \\ 
&+& \left\{\partial_\mu A_\nu^2-\partial_\nu A_\mu^2+g\left[\left(A_\mu^3+E_0t\delta_{\mu 3}\right)A_\nu^1 - A_\mu^1\left(A_\nu^3+E_0t\delta_{\nu 3}\right)\right]\right\}^2\\ 
&+& \left[\partial_\mu A_\nu^3-\partial_\nu A_\mu^3+g\left(A_\mu^1A_\nu^2-A_\mu^2A_\nu^1\right)\right]^2  \nonumber \\
&+& 2E_0^2 -2E_0\left[\partial_\mu A_\nu^3-\partial_\nu A_\mu^3+g\left(A_\mu^1A_\nu^2-A_\mu^2A_\nu^1\right)\right]\left(\delta_{\mu 3}\delta_{\nu 0}-\delta_{\mu 0}\delta_{\nu 3}\right) \nonumber 
\end{eqnarray}

We now expand the above expression to 2$^{nd}$ order in the $A^a _\mu$ which gives
\begin{eqnarray}
	F_{\mu\nu}^a F^{a\mu\nu} &\approx& \left[\partial_\mu A_\nu^1-\partial_\nu A_\mu^1-gE_0 t\left(A_\nu^2\delta_{\mu 3}-A_\mu^2\delta_{\nu 3}\right)\right]^2 \nonumber \\
&+& \left[\partial_\mu A_\nu^2-\partial_\nu A_\mu^2-gE_0 t\left(A_\mu^1\delta_{\nu 3}-A_\nu^1\delta_{\mu 3}\right)\right]^2  \\ 
&+& \left(\partial_\mu A_\nu^3-\partial_\nu A_\mu^3\right)^2+2E_0^2-4E_0\left[\partial_3A_0^3-\partial_0A_3^3+g\left(A_3^1A_0^2-A_3^2A_0^1\right)\right] \nonumber 
\end{eqnarray}
	
We now require that at spatial and temporal infinity the variation of the potential goes to zero 	
$A_0^3\left(x_3\rightarrow\pm\infty\right)=0$ and $A_3^3\left(x_0\rightarrow\pm\infty\right)=0$. This causes the source terms for $A_\mu^3$ in the last term to vanish by partial integration. Note that $\left(\partial_\mu A_\nu^3-\partial_\nu A_\mu^3\right)^2$ is a free Lagrange density, so it is not of interest and can be dropped.
\begin{eqnarray}
	F_{\mu\nu}^a F^{a\mu\nu} &=& \left[\partial_\mu A_\nu^1-\partial_\nu A_\mu^1-gE_0 t\left(A_\nu^2\delta_{\mu 3}-A_\mu^2\delta_{\nu 3}\right)\right]^2 \nonumber \\
&+& \left[\partial_\mu A_\nu^2-\partial_\nu A_\mu^2-gE_0 t\left(A_\mu^1\delta_{\nu 3}-A_\nu^1\delta_{\mu 3}\right)\right]^2 \\
&+& 2E_0^2 - 4 E_0 g\left(A_3^1A_0^2-A_3^2A_0^1\right) \nonumber
\end{eqnarray}
	
We now re-write this using complex potential forms as  
\begin{eqnarray}
F_{\mu\nu}^a F^{a\mu\nu} &=& \left|\partial_\mu\left(A_\nu^1+iA_\nu^2\right)-\partial_\nu\left(A_\mu^1+iA_\mu^2\right)-gE_0t\left[\left(A_\nu^2-iA_\nu^1\right)\delta_{\mu 3}-\left(A_\mu^2-iA_\mu^1\right)\delta_{\nu 3}\right]\right|^2 \nonumber \\ 
&+& 2E_0^2-i2gE_0\left[\left(A_3^1+iA_3^2\right)\left(A_0^1-iA_0^2\right)-\left(A_0^1+iA_0^2\right)\left(A_3^1-iA_3^2\right)\right]
\end{eqnarray}
	
We now replace $A_\mu^1$ and $A_\mu^2$ by defining the following complex potetnials
\begin{equation}
W_\mu=\frac{1}{\sqrt{2}}\left(A_\mu^1+iA_\mu^2\right) \hspace{10pt} , \hspace{10pt} 
W_\mu^\dagger=\frac{1}{\sqrt{2}}\left(A_\mu^1-iA_\mu^2\right).
\end{equation}
	
In terms of these new, complex potentials $F_{\mu\nu}^a F^{a\mu\nu}$ becomes  
\begin{eqnarray}
F_{\mu\nu}^a F^{a\mu\nu} &=& \left|\sqrt{2}\partial_\mu W_\nu-\sqrt{2}\partial_\nu W_\mu-i\sqrt{2}gE_0 tW_\nu\delta_{\mu 3}+i\sqrt{2}gE_0t W_\mu\delta_{\nu 3}\right|^2 \nonumber \\
&+& 2E_0^2-i2gE_0\left(2W_3^\dagger W_0-2W_0^\dagger W_3\right)
\end{eqnarray}
This finally leads to the following Lagrange density for gluons in the background potential of a uniform color electric field 
\begin{eqnarray}
\mathcal{L} &=& -\frac{1}{4}F_{\mu\nu}^a F^{a\mu\nu} \\
&=& -\frac{1}{2}\left|\left(\partial_\mu-igE_0t\delta_{\mu 3}\right)W_\nu-\left(\partial_\nu -igE_0t\delta_{\nu 3}\right)W_\mu\right|^2-igE_0\left(W_3^\dagger W_0-2W_0^\dagger W_3\right)-\frac{1}{2}E_0^2 \nonumber
\end{eqnarray}
	
This Lagrange density leads to the following equation of motion 
\begin{equation}
\label{eqn-mot}
\left(\partial^\mu - igE_0t\delta^{\mu 3}\right)\left[\left(\partial_\mu-igE_0t\delta_{\mu 3}\right)W_\nu-\left(\partial_\nu-igE_0t\delta_{\nu 3}\right)W_\mu\right]-igE_0\left(\delta_{\nu 3}W_0-\delta_{\nu 0}W_3\right)=0.
\end{equation}
	
Choosing the background gauge condition $\left(\partial^\mu-igE_0t\delta^{\mu 3}\right)W_\mu=0$ \footnote{This background gauge condition singles out the physical degrees of freedom.
Formally one can show this by using the gauge condition to find the corresponding ghost fields and showing that these cancel the contribution of $A^3 _\mu$ and those components of
$W_\mu$ for which $\left(\partial^\mu-igE_0t\delta^{\mu 3}\right)W_\mu \ne 0$. Thus although ghost fields do not appear explicitly in our calculations they are taken into account
implicitly via the background gauge condition and our taking $A_0^3\left(x_3\rightarrow\pm\infty\right)=0$ and $A_3^3\left(x_0\rightarrow\pm\infty\right)=0$.} simplifies (\ref{eqn-mot}) to

\begin{equation}	
\label{eqn-mot2}
\left(\partial^\mu-igE_0t\delta^{\mu 3}\right)\left(\partial_\mu-igE_0t\delta_{\mu 3}\right)W_\nu=
-2igE_0\left(W_0\delta_{\nu 3}-W_3\delta_{\nu 0}\right)
\end{equation}

The above equation can be written in matrix form as
	
\begin{eqnarray}	
\label{eqn-mot3}
\left(\partial^\mu-igE_0t\delta^{\mu 3}\right)^2W_\nu-2gE_0 
\left[ \begin{array}{cccc}
		0 & 0 & 0 & -i\\
		0 & 0 & 0 & 0\\
		0 & 0 & 0 & 0\\
		i & 0 & 0 & 0 
\end{array} \right]
		W_\mu=0
\end{eqnarray}			
		
The four eigenvalues for this matrix are $\pm 1$ and a double eigenvalue of zero. The zero eigenvalues are excluded by our chosen gauge condition so for now we consider just the eigenvalues $\pm 1$. Thus 
(\ref{eqn-mot3}) becomes

\begin{equation}
\label{eqn-mot4}
\left[\left(\partial^\mu-igE_0t\delta^{\mu 3}\right)^2 W_\nu-2gE_0\left(\pm 1\right)\right]W_\nu=0.
\end{equation}

Recalling that we are using the metric signature $(+,-,-,-)$ (\ref{eqn-mot4}) can be expanded as
\begin{equation}
\label{eqn-mot5}
\left\{\pderiv{^2}{t^2}-\pderiv{^2}{x_1^2}-\pderiv{^2}{x_2^2}+\left(i\pderiv{}{x_3}+gE_0t\right)^2 \mp 2gE_0\right\}W_\nu=0
\end{equation}

Next we Fourier transform (\ref{eqn-mot5}) to $\tilde{W}_\nu(t,k_1,k_2,k_3)$ with the result

\begin{equation}
\label{eqn-mot5a}
\left\{\pderiv{^2}{t^2}+k_1^2+k_2^2+g^2E_0^2\left(\frac{k_3}{gE_0}+t\right)^2 \pm 2gE_0\right\}\tilde{W}_\nu=0
\end{equation}

Now making the substitution $\displaystyle t'=\frac{k_3}{gE_0}+t$ (for which one has $\pderiv{}{t}=\pderiv{}{t'}$) and then 
performing a rotaion to imaginary time, $t' \rightarrow -i\tau$, $E_0 \rightarrow -iE_0$ we arrive at

\begin{equation}
\label{eqn-mot6}
\left\{-\pderiv{^2}{\tau^2}+g^2E_0^2\tau'^2+k_1^2+k_2^2 \pm 2igE_0\right\}\tilde{W}_\nu=0
\end{equation}
	
The first two terms in (\ref{eqn-mot6}) correspond to a harmonic oscillator with frequency $\omega = g E_0$, which has eigenvalues given
by $\displaystyle -\pderiv{}{\tau^2}+g^2E_0^2\tau^2 \rightarrow 2\left(n+\frac{1}{2}\right)gE_0$. With this (\ref{eqn-mot6}) becomes
$\left\{\left(2n+1\right)gE_0+k_1^2+k_2^2 \pm igE_0\right\}\tilde{W}_\nu=0$. From this equation we can read of the eigenvalues as	
\begin{equation}
\label{evalue}
\lambda_n=\left(2n+1\right)gE_0+k_1^2+k_2^2 \pm 2igE_0.
\end{equation}
Note that $\lambda_n$ is a combination of discrete ({\it i.e.} $\left(2n+1\right)gE_0$) and continuous ({\it i.e.} $k_1^2+k_2^2$) parts. 
Substituting these eigenvalues from (\ref{evalue}) into (\ref{vac-vac2}) we find that $\zeta$ becomes
\begin{equation}
\label{zeta-2}
\zeta=-\sum\limits_n\int\limits_0^\infty\frac{\mathrm{d}s}{s}\exp\left\{-\left[\left(2n+1\right)gE_0+k_1^2+k_2^2 \pm 2igE_0\right]s\right\}
\end{equation}
	
We take our system to be quantized in a cubical spatial volume with sides of length $L$ and over a total (imaginary) interaction time 
$\tau=-iT$ which then turns (\ref{zeta-2}) into
\begin{equation}
\label{zeta-3}
\zeta=-L\int\limits_{-\infty}^\infty\frac{\mathrm{d}k_1}{(2\pi)}L\int\limits_{-\infty}^\infty\frac{\mathrm{d}k_2}{(2\pi)}L\int\limits_0^{gE_0\tau}\frac{\mathrm{d}k_3}{(2\pi)}\sum\limits_{n=0}^\infty\int\limits_0^\infty\frac{\mathrm{d}s}{s}\exp\left\{-\left[\left(2n+1\right)gE_0+k_1^2+k_2^2 \pm 2igE_0\right]s\right\}.
\end{equation}
	
The $k_1$ and $k_2$ integrations related to the momentum in the free directions and are simple Gaussian integrals, which give two factors of $\displaystyle \sqrt{\frac{\pi}{s}}$. 	
The integration in the $k_3$ direction is related to the momentum in the $z$ direction which is the direction of the chromoelectric field. Thus as in \cite{holstein}
this integration is constrained to the range $0<k_3<gE_0 \tau$. Performing the $k_1, k_2, k_3$ integrations gives
\begin{equation}
\label{zeta-4}
\zeta=-L^3\tau\frac{gE_0}{8\pi^2}\sum\limits_{n=0}^\infty\int\limits_0^\infty\frac{\mathrm{d}s}{s^2}\exp\left\{-\left[\left(2n+1\right)gE_0 \pm 2igE_0\right]s\right\}
\end{equation}

We note that $\displaystyle \sum\limits_{n=0}^\infty e^{-\left(2n+1\right)gE_0s}=\frac{1}{2}\frac{1}{\sinh(gE_0s)}$ which then transforms (\ref{zeta-4}) into
\begin{equation}
\label{zeta-5}
\zeta=-L^3\tau\frac{gE_0}{16\pi^2}\int\limits_0^\infty\frac{\mathrm{d}s}{s^2}\frac{e^{\pm 2igE_0s}}{\sinh(gE_0s)}
\end{equation}

We now return to real time via the rotation $\tau \rightarrow iT$ which also involves changing the magnitude of the chromoelectric field as $E_0 \rightarrow iE_0$. This gives
\begin{equation}
\label{zeta-6}
\zeta=-iL^3 T\frac{gE_0}{16\pi^2}\int\limits_0^\infty\frac{\mathrm{d}s}{s^2}\frac{e^{\mp 2gE_0s}}{\sin(gE_0s)}
\end{equation}

Equation (\ref{zeta-6}) is now at the point in the electromagnetic calculation given by equation (\ref{zeta}), but now the exponential factor involving the
electron mass ($\exp[-m^2 s]$) is replaced by an exponential suppression involving the field strength ($e^{\mp 2gE_0s}$). As before if $\zeta$ has a real 
part ({\it i.e.} if the integral in (\ref{zeta-6}) has an imaginary contribution) there will be particle production. As in the case of the electromagnetic integral
in (\ref{zeta}) the integral in (\ref{zeta-6}) does have an imaginary contribution coming from the poles of $\frac{1}{\sin(gE_0s)}$ whihc are located at $\displaystyle s_n=\frac{n\pi}{gE_0}$, 
where $n$ is an integer. As before we ignore the singularity at $s=0$. The integration contours are infinitesimal semicircular in the upper half plane and from (\ref{zeta-6})
this gives
\begin{eqnarray}
\label{gamma-2}
\gamma &=& {\rm Re}(\zeta)=-iL^3T\frac{gE_0}{16\pi^2}\sum_{n=1}^\infty\int\limits_{s_n-\epsilon}^{s_n+\epsilon}\frac{\mathrm{d}s}{s^2}\frac{e^{\mp 2gE_0s}}{\sin(gE_0s)} \nonumber \\
&=& -iL^3T\frac{gE_0}{16\pi^2}\sum_{n=1}^\infty\int\limits_{s_n-\epsilon}^{s_n+\epsilon}\frac{\mathrm{d}s}{s^2}\frac{e^{\mp 2gE_0s}}{\cos(gE_0s_n)gE_0\left(s-s_n\right)} ~. 
\end{eqnarray}
In the last step we have expanded $\sin(gE_0s)$ around the poles at $s_n$. From the Residue theorem, the poles of the integral in (\ref{gamma-2}) give  
$-i \pi  \times \sum {\rm Res}({\rm function})$, where the sum is over the residue of the integrand. The result is 
\begin{equation}
\label{gamma-3}
\gamma= {\rm Re}(\zeta)=-L^3T\frac{1}{16\pi}\sum_{n=1}^\infty\frac{e^{\mp 2gE_0s_n}}{s_n^2\cos(gE_0s_n)}
= L^3T\frac{g^2E_0^2}{16\pi^3}\sum\limits_{n=1}^\infty (-1)^{n+1}\frac{e^{\mp 2n\pi}}{n^2} ~,
\end{equation}
where in the last step we have substitute in the poles $s_n=\frac{n\pi}{gE_0}$ and simplify. The $(-1)^{n+1}$ comes from $-\cos(n\pi)$.
The $e^{+ n\pi}$ choice in (\ref{gamma-3}) leads to a divergent $\gamma$ so we
take the $e^{- n\pi}$ choice in (\ref{gamma-3}) which leads to 
\begin{equation}
\label{gamma-4}
\gamma =L^3T\frac{g^2E_0^2}{16\pi^3}\sum\limits_{n=1}^\infty (-1)^{n+1}\frac{e^{-2n\pi}}{n^2} \approx 0.00187 \times L^3T\frac{g^2E_0^2}{16\pi^3} ~.
\end{equation}
In the last step we have carried out the sum numerically with the result $\sum\limits_{n=1}^\infty (-1)^{n+1}\frac{e^{-2n\pi}}{n^2} \approx 0.00187$. 

Calculations similar to the above were carried out in \cite{nayak1, nayak2, nayak3} for gluon production 
in a constant $SU(3)$ color electric field. In these works the integration over the transverse momentum was not carried out, but if one
does carry out the integration of the transverse momentum of the results in \cite{nayak1, nayak2, nayak3} one finds that our result 
for $\gamma$ in (\ref{gamma-4}) is consistent with these previous results. In particular there is no mass suppression (as expected) and the
production rate is proportional to $g^2E_0^2$

\section{Uniform Color Magnetic Field Calculation}

In this appendix we show that, unlike the uniform color electric field of appendix A, a uniform color magnetic field does not produce gluons.
This is similar to what occurs in the electromagnetic Schwinger effect -- a uniform magnetic field does not produce electrons/positrons.  

A constant color magnetic field in the $\hat{z} = 3$ spatial direction and in the $a=3$ color direction, 
\begin{equation}
\label{bqcd}
B^a _\mu \rightarrow B^{3} _{3} = F^3 _{12} = \partial _1 A_2 ^3 - \partial _2 A_1 ^3 +g \epsilon^{3bc} A^b_1 A^c_2 = H_0 \hat{z} ~,
\end{equation}
can be obtained by the potential 
\begin{equation}
\label{aqcd-1}
A_\mu^{a} = - H_0 x_1 \delta_{\mu 2} \delta^{a3} ~.
\end{equation}
With the potential in (\ref{aqcd-1}) we perform a long calculation that is similar to the one for the constant color electric field, which leads to 
\begin{eqnarray}	
\label{eqn-mot-b}
\left(\partial^\mu-igH_0x\delta^{\mu 2}\right)^2W_\nu-2gH_0
\left[ 
\begin{array}{cccc}
		0 & 0 & 0 & 0\\
		0 & 0 & i & 0\\
		0 & -i & 0 & 0\\
		0 & 0 & 0 & 0
\end{array} \right]		
		W_\mu=0 ~.
\end{eqnarray}	
which is the color magnetic version of (\ref{eqn-mot3}). Expanding (\ref{eqn-mot-b}) out we arrive at the color magnetic version (\ref{eqn-mot5})
which leads to 
\begin{equation}
\label{eqn-mot-b1}
\left[\pderiv{^2}{t^2} -\pderiv{^2}{x^2} -\pderiv{^2}{z^2}  \mp 2gH_0 - \left( i\pderiv{}{y}+gH_0 x \right)^2\right] {W}_\nu=0
\end{equation}
Next we Fourier transform (\ref{eqn-mot-b1}) to $\tilde{W}_\nu(t,k_1,k_2,k_3)$ gives the color magnetic version of (\ref{eqn-mot5a})
\begin{equation}
\label{eqn-mot-b2}
\left[-E^2+k_3^2 \mp 2gH_0-\pderiv{^2}{x^2}+g^2H_0^2\left(x+\frac{k_2}{gH_0}\right)^2\right]\tilde{W}_\nu=0 ~.
\end{equation}
Making the change of variable $\eta=x+\frac{k_2}{gH_0}$ this becomes
\begin{equation}
\label{eqn-mot-b3}
\left[-E^2+k_3^2 \mp 2gH_0-\pderiv{^2}{\eta^2}+g^2H_0^2\eta^2\right]\tilde{W}_\nu=0 ~.
\end{equation}
The $-\pderiv{^2}{\eta^2}+g^2H^2\eta^2$ part of the above equation is the equation for a simple harmonic oscillator 
which has eignevalues $2\left(n+\frac{1}{2}\right)gH_0$. Substituting this eigenvalue into (\ref{eqn-mot-b3})
leads to 
\begin{equation}
\label{eqn-mot-b4}
\left[-E^2+k_3^2 \mp 2gH_0+2\left(n+\frac{1}{2}\right)gH_0\right]\tilde{W}_\nu=0
\end{equation}
We now rotate to imaginary time (Euclidean) as $t \rightarrow -i\tau$ and in conjunction with this
we rotate the energy as $E \rightarrow =-iE$. This turns (\ref{eqn-mot-b4}) into
\begin{equation}
\label{eqn-mot-b5}
\left[E^2+k_3^2 \mp 2gH_0+2\left(n+\frac{1}{2}\right)gH_0\right]\tilde{W}_\nu=0
\end{equation}
The eigenvalues for the system can then be written out as
\begin{equation}
\label{eqn-mot-b6}
\lambda_n=E^2+k_3^2 \mp 2gH_0+2\left(n+\frac{1}{2}\right)gH_0
\end{equation}
Now substituting (\ref{eqn-mot-b6}) into (\ref{vac-vac2}) yields $\zeta$ for this case as 
\begin{equation}
\label{eqn-mot-b7}
\zeta=-\sum\limits_n\int_0^\infty \frac{\mathrm{d}s}{s}\exp\left\{-\left[E^2+k_3^2+(2n+1)gH_0 \mp 2gH_0\right]s\right\}
\end{equation}
Recalling that we are quantizing in a cube with sides length $L$ and over a total (imaginary) interaction time $\tau=-iT$ transforms
(\ref{eqn-mot-b7}) into 
\footnote{From equation 8.25 in \cite{greiner} the integral over the momenta in the $x$ and $y$ directions is
$\displaystyle \int\frac{\mathrm{d}k_1\mathrm{d}k_2}{(2\pi)^2} \rightarrow \frac{\int k\mathrm{d}k}{2\pi} \rightarrow \frac{gH_0}{2\pi}\sum\limits_n$}
\begin{equation}
\label{eqn-mot-b8}	
\zeta=-\tau\int\limits_{-\infty}^\infty \frac{\mathrm{d}E}{(2\pi)} L\int\limits_{-\infty}^\infty\frac{\mathrm{d}k_3}{(2\pi)} L^2\frac{gH_0}{2\pi}\sum\limits_{n=0}^\infty\int\limits_0^\infty\frac{\mathrm{d}s}{s}\exp\left\{-\left[E^2+k_3^2+(2n+1)gH_0 \mp 2gH_0\right]s\right\} ~.
\end{equation}
The $E$ and $k_3$ integrations are simple Gaussian integrals that give two factors of $\sqrt{\frac{\pi}{s}}$. Thus
(\ref{eqn-mot-b8}) simplifies to 
\begin{equation}
\label{eqn-mot-b9}
\zeta=-\tau L^3\frac{gH_0}{8\pi^2}\sum\limits_{n=0}^\infty\int\limits_0^\infty\frac{\mathrm{d}s}{s^2}\exp\left\{-\left[(2n+1)gH_0 \mp 2gH_0\right]s\right\} ~.
\end{equation}
We now use the geometric series  $\displaystyle \sum\limits_{n=0}^\infty e^{-(2n+1)gHs}=\frac{1}{2}\frac{1}{\sinh(gHs)}$ to write 
(\ref{eqn-mot-b9}) as
\begin{equation}
\label{eqn-mot-b10}
\zeta=-\tau L^3\frac{gH_0}{16\pi^2}\int\limits_0^\infty \frac{\mathrm{d}s}{s^2}\frac{e^{\pm 2gH_0s}}{\sinh(gH_0s)}
\end{equation}
We now return to real time via the rotation $\tau \rightarrow iT$.
\begin{equation}
\label{eqn-mot-b11}
\zeta=-iTL^3\frac{gH_0}{16\pi^2}\int\limits_0^\infty \frac{\mathrm{d}s}{s^2}\frac{e^{\pm 2gH_0s}}{\sinh(gH_0s)}
\end{equation}
In contrast to the color electric case of appendix A where the rotation back to real time, $\tau \rightarrow iT$,
was accompanied by a rotation of the electric field amplitude, $E_0 \rightarrow i E_0$, here the rotation back to real time does
not lead to a change in the color magnetic field amplitude. In comparing (\ref{eqn-mot-b11}) with (\ref{zeta-6}) one finds that 
(\ref{eqn-mot-b11}) does not have the series of poles at $s_n = n \pi / q E_0$ that are found in (\ref{zeta-6}). Therefore the
integral in (\ref{eqn-mot-b11}) does not have any imaginary contributions coming from the residue theorem was was the case 
for the integral in (\ref{zeta-6}). Thus from (\ref{eqn-mot-b11}) the real part of $\zeta$ is zero so $Re (\zeta) = \gamma = 0$. 
Thus in the case of a uniform color magnetic field there is no particle production as is expected form the electromagnetic case 
where one finds that a constant magnetic field does not produce electron/positron pairs.

\end{document}